\documentclass[preprint,12pt,authoryear]{elsarticle}

\usepackage{booktabs}
\usepackage{makecell}
\usepackage{framed, color}
\usepackage{soul}
\usepackage{subcaption}% <-- added
\usepackage{bbm}
\usepackage{dsfont}
\usepackage{amsmath,amssymb}
\usepackage{url}
\usepackage{hyperref}

\newcommand{\eg}{\text{e.g.}, }
\newcommand{\ie}{\text{i.e.}, }
\newcommand{\etal}{\text{et al.}~}
\newcommand{\vs}{\text{vs.} }

\newcommand{\one}[0] {{(1)}~}
\newcommand{\two}[0] {{(2)}~}
\newcommand{\three}[0] {{(3)}~}
\newcommand{\four}[0] {{(4)}~}

\journal{}

\begin{document}

\begin{frontmatter}

\title{Majority Voting of Doctors Improves Appropriateness of AI Reliance in Pathology}

\author[1]{Hongyan Gu}
\author[1]{Chunxu Yang}
\author[2]{Shino Magaki}
\author[3]{Neda Zarrin-Khameh}
\author[4]{Nelli S. Lakis}
\author[5]{Inma Cobos}
\author[2]{Negar Khanlou}
\author[6]{Xinhai R. Zhang}
\author[4]{Jasmeet Assi}
\author[7]{Joshua T. Byers}
\author[4]{Ameer Hamza}
\author[8]{Karam Han}
\author[4]{Anders Meyer}
\author[2]{Hilda Mirbaha}
\author[9]{Carrie A. Mohila}
\author[4]{Todd M. Stevens}
\author[10]{Sara L. Stone}
\author[1]{Wenzhong Yan}
\author[4]{Mohammad Haeri}
\author[1]{Xiang `Anthony' Chen}

\address[1]{Department of Electrical and Computer Engineering, University of California, Los Angeles, Los Angeles, USA}
\address[2]{Department of Pathology and Laboratory Medicine, UCLA David Geffen School of Medicine, Los Angeles, USA}
\address[3]{Department of Pathology and Laboratory Medicine, Baylor College of Medicine and Ben Taub Hospital, Houston, USA}
\address[4]{Department of Pathology and Laboratory Medicine, University of Kansas Medical Center, Kansas City, USA}
\address[5]{Department of Pathology, Stanford School of Medicine, Stanford, USA}
\address[6]{McGovern Medical School, University of Texas Health Science Center at Houston, Houston, USA}
\address[7]{School of Medicine, University of California, San Francisco, San Francisco, USA}
\address[8]{Department of Pathology and Laboratory Medicine, University of Wisconsin-Madison, Madison, USA}
\address[9]{Department of Pathology, Baylor College of Medicine and Texas Children's Hospital, Houston, USA}
\address[10]{Department of Pathology, Hospital of the University of Pennsylvania, Philadelphia, USA}

\cortext[cor1]{Corresponding authors: Hongyan Gu (ghy@ucla.edu), Mohammad Haeri (mhaeri@kumc.edu)}

\begin{abstract}
As Artificial Intelligence (AI) making advancements in medical decision-making, there is a growing need to ensure doctors develop appropriate reliance on AI to avoid adverse outcomes. However, existing methods in enabling appropriate AI reliance might encounter challenges while being applied in the medical domain. With this regard, this work employs and provides the validation of an alternative approach -- majority voting -- to facilitate appropriate reliance on AI in medical decision-making. This is achieved by a multi-institutional user study involving 32 medical professionals with various backgrounds, focusing on the pathology task of visually detecting a pattern, mitoses, in tumor images. Here, the majority voting process was conducted by synthesizing decisions under AI assistance from a group of pathology doctors (pathologists). Two metrics were used to evaluate the appropriateness of AI reliance: Relative AI Reliance (RAIR) and Relative Self-Reliance (RSR). Results showed that even with groups of three pathologists, majority-voted decisions significantly increased both RAIR and RSR -- by approximately 9\% and 31\%, respectively -- compared to decisions made by one pathologist collaborating with AI. This increased appropriateness resulted in better precision and recall in the detection of mitoses. While our study is centered on pathology, we believe these insights can be extended to general high-stakes decision-making processes involving similar visual tasks.
\end{abstract}

\begin{highlights}
\item Multi-institutional user study with medical doctors from diverse backgrounds
\item Two quantifiable metrics for evaluating appropriateness of AI reliance
\item Majority voting improves appropriateness of AI reliance in pathology decisions
\item Increased appropriateness elevates precision and recall
\item Grouped pathologists improve qualities in visual decision-making
\end{highlights}


\end{frontmatter}

% Main text

\section{Introduction}

Although J.C.R. Licklider introduced the concept of `man-computer symbiosis' in 1960 \citep{licklider1960man}, it was not until the last decade that this vision became a more promising reality \citep{doi:10.1126/science.aaa8415}. By 2023, Artificial Intelligence (AI) has been increasingly discussed to augment humans in critical tasks \citep{bi2019artificial, surden2019artificial, grigorescu2020survey}. Especially in the medical domain of pathology, AI has been showcased to increase doctors' accuracy and speed \citep{litjens2016deep, 10.1145/3397481.3450681, van2021deep, ba2022assessment}, consistency \citep{balkenhol2019deep, van_bergeijk_deep_2023}, and confidence \citep{10.1145/3577011}. However, because pathology AI was often trained from a limited dataset its performance varied while being applied to data from new patients and hospitals \citep{stacke2020measuring, aubreville2021quantifying, 10.1145/3449084}. As such, it is critical for pathologists to develop appropriate reliance while collaborating with AI, \ie to appropriately accept correct AI recommendations and reject the wrong ones.

Although there is a lack of data in pathology, research in the general domain has explored methodologies to develop appropriate reliance, focusing on reducing humans' over-reliance on AI (\ie enhancing humans' ability to reject wrong AI recommendations). Strategies, including the cognitive forcing function \citep{10.1145/3449287} and altering the interaction speed \citep{10.1145/3359204, 10.1145/3512930, 10.1007/978-3-031-35891-3}, have shown promising results. Additionally, effective onboarding \citep{passi2022overreliance} and improving AI literacy \citep{10.1145/3313831.3376727} were recommended and can be achieved by informing users of AI details \citep{10.1145/3359206, 10.1145/3411764.3445385, jacobs_how_2021}. However, incorporating these methods into routine medical practice presents challenges: Cognitive forcing functions could drive medical practitioners to develop algorithm aversion, leading them to reject AI recommendations even when they were correct \citep{efendic2020slow, 10.1145/3531146.3533193}. Moreover, previous studies have reported that the improvements in task accuracy with enhanced AI literacy were marginal \citep{10.1145/3313831.3376873, leichtmann2023effects}.

Another popular approach aims to employ explainable AI (XAI) to reduce over-reliance \citep{7349687, 10.1145/3313831.3376873, 10.1145/3351095.3372852, 10.1145/3411764.3445717}. However, the efficacy of XAI is countered in part by the cognitive effort for understanding these explanations \citep{10.1145/3579605}. Adding XAI-related content might increase doctors' cognitive burden, possibly causing them to overlook XAI. Therefore, there remains a pressing need for alternative strategies to foster appropriate AI reliance in medical applications.

By reviewing pathologists' decision-making workflows, we found that the critical decisions were usually determined through a combined judgment among multiple doctors \citep{black1999consensus}. The underlying intuition was that a group of pathologists might produce safer and more rational judgments while working together \citep{black1999consensus}. In the context of AI, recent studies have employed majority voting among pathologists' AI-assisted decisions to collect annotations for datasets \citep{bertram_large-scale_2019, aubreville_completely_2020}. However, there is a lack of empirical evidence supporting that such a majority voting approach would enable appropriate reliance.

This research aims to provide the validation of the majority voting on enabling the appropriate AI reliance in pathology decision-making, with a focus on a visual search task of detecting ``mitosis,'' a  critical histology pattern for tumor grading \citep{collan_standardized_1996, meyer_breast_2005}. 32 medical professionals in pathology from ten institutions participated in a multi-stage user study, where they detected mitoses manually, first, and with AI assistance after a wash-out period. Here, the majority voting decisions were synthesized according to the AI-assisted decisions from an odd number of randomly-selected pathologist participants. Two metrics were employed to measure the appropriateness of AI reliance: ``relative AI reliance'' and ``relative self-reliance'' \citep{10.1145/3581641.3584066}. The result showed that the majority voting decisions from as few as three pathologists showed significantly higher relative AI reliance ($\sim 9\%$ increase) and relative self-reliance ($\sim 31\%$ increase), compared to one pathologist collaborating with AI, respectively. The precision and recall of majority voting decisions also increased: Those from three AI-assisted pathologists could achieve a mean precision of 0.902 and a recall of 0.843. As a comparison, the mean precision and recall for one-pathologist-AI collaboration were 0.824 and 0.817, respectively. Furthermore, the majority voting decisions could also have a higher chance of achieving super-AI performance in the recall.

\subsection{Contributions}
This research showcases that majority voting can enable appropriate AI reliance for pathology decision-making. Throughout a multi-institutional study amongst 32 pathology professionals, this research presents the effectiveness of majority voting in a high-stakes medical task, which can ultimately benefit patient management. This signifies a transformation from the traditional one-human-AI collaboration to harnessing group decision-makings of AI-assisted medical professionals. While our primary focus has been on pathology, we envision that the insights of this study can have broader implications for leveraging collective human-AI decision-making in other high-stakes visual search tasks, such as detecting explosives from X-ray scans or disaster assessment from satellite imagery for emergency response efforts.

\section{Related Work}

\subsection{Enabling Appropriate AI Reliance}

According to \cite{passi2022overreliance, 10.1145/3579605, 10.1145/3581641.3584066}, two goals should be achieved to enable appropriate AI reliance: \one mitigating over-reliance, where humans can identify and reject AI's incorrect recommendations, and \two reducing under-reliance, where humans can overcome their aversion of AI and accept its correct recommendations.

In the context of enabling appropriate AI reliance, this is a tendency in research to study mechanisms and counter-measures for over-reliance. For instance, the cognitive forcing function, which prompts users to think analytically before decision-making, has shown promise \citep{10.1145/3449287}. Similarly, altering the interaction speed, where enlonging the AI response time, can instigate users' reflective thinking. Therefore, over-reliance incidents could be reduced \citep{10.1145/3359204, 10.1145/3512930, 10.1007/978-3-031-35891-3}. Other approaches aim to enhance users' onboarding process, such as improving AI literacy \citep{10.1145/3313831.3376873, 10.1145/3313831.3376727, leichtmann2023effects}, where users are informed of AI details \citep{10.1145/3359206, 10.1145/3411764.3445385, jacobs_how_2021}. However, translating these approaches to the medical domain may encounter two challenges. Firstly, introducing cognitive forcing functions or altering interaction speed could develop `algorithm aversion,' especially when medical tasks are time-sensitive \citep{efendic2020slow, 10.1145/3531146.3533193}. Secondly, the efficacy of enhancing AI literacy also appeared marginal, possibly because of the difficulties in educating users within a limited timeframe \citep{10.1145/3313831.3376873, leichtmann2023effects}.

Besides these, another popular approach is XAI, aiming to reduce over-reliance by enabling users to understand AI's reasoning \citep{7349687, 10.1145/3313831.3376873, 10.1145/3351095.3372852, 10.1145/3411764.3445717}. Nonetheless, numerous studies have failed to observe the anticipated effectiveness of XAI \citep{10.1145/3514094.3534128}: The potential benefits of XAI may be offset by the cognitive efforts of interpreting them \citep{10.1145/3579605}. Given the already high cognitive demands of medical professionals, this might result in XAI being less referred to, countering its potential benefits. This issue of appropriateness usage of XAI in medicine was raised by \cite{holzinger2019causability}. Further research suggested causability, an ability of an explanation that can enable casual understanding of medical experts, should also considered and measured to achieve better efficiency, effectiveness, and user satisfaction \citep{holzinger2020measuring, plass_explainability_2023}.

Notably, most of the research mentioned above focuses on scenarios where one human collaborates with AI. Different from these studies, our approach learns from how critical pathology decisions are usually made -- through a group of pathologists \citep{black1999consensus}. Specifically, we employ a majority voting approach to synthesize decisions from multiple AI-assisted pathologists. Based on the results, the majority voting approach can effectively enable appropriate AI reliance, which sheds light on the potential of involving multiple professionals in critical decision-making.

\subsection{Decision-Making Processes by Multiple Medical Professionals}

Different from one medical professional examining the specimens \citep{pohn_towards_2019, pena2009does}, decision-making processes by medical professionals require communication, discussion, and result sharing \citep{murphy1998consensus, black1999consensus}. Nowadays, there are three primary approaches: \one the Delphi method, \two the nominal group technique, and \three the consensus development conference. These methods have been historically utilized in medical decision-making and guideline formulation \citep{murphy1998consensus, black1999consensus}.
 
The Delphi method follows an iterative process: Each group member makes a decision first anonymously. Next, their opinions are collected, summarized, and then sent back to all members. Upon reviewing this summary, each member may choose to modify their opinions secretly. This process may be iterated multiple times to resolve potential conflicts \citep{taze2022developing}.
The nominal group technique \citep{van1972nominal} also starts by collecting each member's opinions. Then, in a structured face-to-face meeting, these opinions are presented and discussed. Next, each member ranks the presented opinions according to their preferences. These rankings will be summed and posted for further discussion \citep{mcmillan2016use}.
The consensus development conference \citep{ferguson1996nih} is more open in its structure. Group members are presented with evidence by external experts during a series of face-to-face meetings. Group members can then question the expert presenters, and attempt to reach an agreement afterward.

Regarding employing multiple pathologists to make decisions with AI assistance, recent research has implemented majority voting among three doctors to label data for AI development \citep{bertram_large-scale_2019, aubreville_completely_2020}. This process involves two pathologists independently annotating data with AI assistance. If there were conflicts, a third pathologist joined and annotated again to formulate majority \citep{MONTEZUMA2023100086}. However, these studies primarily focus on data labeling for AI development, often featuring non-diverse participant pools (typically three pathologists) with similar backgrounds. Furthermore, they lack analyses of AI reliance metrics, which is critical for assessing the method's quality.

Our study aims to fill the gap by providing a comprehensive experiment and evaluation of the majority voting on pathology decision-making. To achieve this, we hosted a multi-stage, multi-institutional user study involving 32 medical professionals in pathology with varied experience levels. We examined the quality of these majority-voted, AI-assisted decisions from two angles: reliance on AI and correctness. Additionally, we evaluated the potential costs of employing this method, providing empirical data for future research in HCI, cognitive science, and medicine.

\subsection{Human-AI Collaboration in Pathology}
AI, particularly deep learning, holds promise in performing a wide variety of pathology tasks to assist medical professionals \citep{regitnig_expectations_2020}, ranging from conducting high-level diagnoses (\eg prostate cancer grading \citep{pantanowitz2020artificial}), to detecting low-level pathological patterns (\eg cell detection \citep{amgad2022nucls}). Notably, recent studies have suggested that AI's performance is on par with human experts in specific pathology tasks \citep{HEKLER201979, zhang2019, wang2023}. Due to legislative and ethical concerns, AI algorithms and software cannot replace pathologists' examinations \citep{chauhan2021ethics, veale2021demystifying}. Instead, they are regarded as medical devices to assist doctors\footnote{\url{https://www.fda.gov/media/145022/download}}, with one designed for prostate cancer pathology receiving the first official approval in 2021\footnote{\url{https://www.fda.gov/news-events/press-announcements/fda-authorizes-software-can-help-identify-prostate-cancer}}. By 2023, a plethora of human-AI collaborative tools have been introduced, demostrating improvements in speed and correctness in detecting pathological patterns \citep{litjens2016deep, 10.1145/3397481.3450681, van2021deep, ba2022assessment}, inter-observer consistencies \citep{balkenhol2019deep, van_bergeijk_deep_2023}, mental workload and confidence \citep{10.1145/3577011, 10.1145/3544548.3580694}, compared to pathologists examining manually.

Among these improvements, of particular interest is achieving complementary team performance,  where pathologist-AI collaboration could outperform both the pathologist and AI \citep{10.1145/3411764.3445717}: In 2016, Wang \etal~reported that combining AI and pathologist's predictions could reduce error rates in breast cancer classification, theoretically confirming the existence of such complementary team performance \citep{wang2016deep}. Despite this theoretical backing, there is a lack of empirical evidence to support it for the pathology domain. In the general domain, several studies have failed to observe the task accuracy improvement in human-AI collaboration compared to AI alone \citep{10.1145/3411764.3445717, 10.1145/3287560.3287590, jacobs_how_2021}. This issue may stem from users' accepting incorrect AI recommendations, a factor that can significantly impact the outcome of human-AI collaboration \citep{10.1145/3313831.3376219, 10.1145/3555572, 10.1145/3579605}.

To date, research remains sparse on how pathologists would rely on AI. This work fills this gap and by recruiting pathology professionals and studying their AI reliance, which can help future researchers understand pathologists' behavior, and develop potential solutions to enable appropriate AI reliance. 

% Medical Background
\section{Task Design \& Medical Background}

\subsection{Task Selection \& Generalizability of the Task}

\begin{figure}
    \centering
    \includegraphics[width=0.8\linewidth]{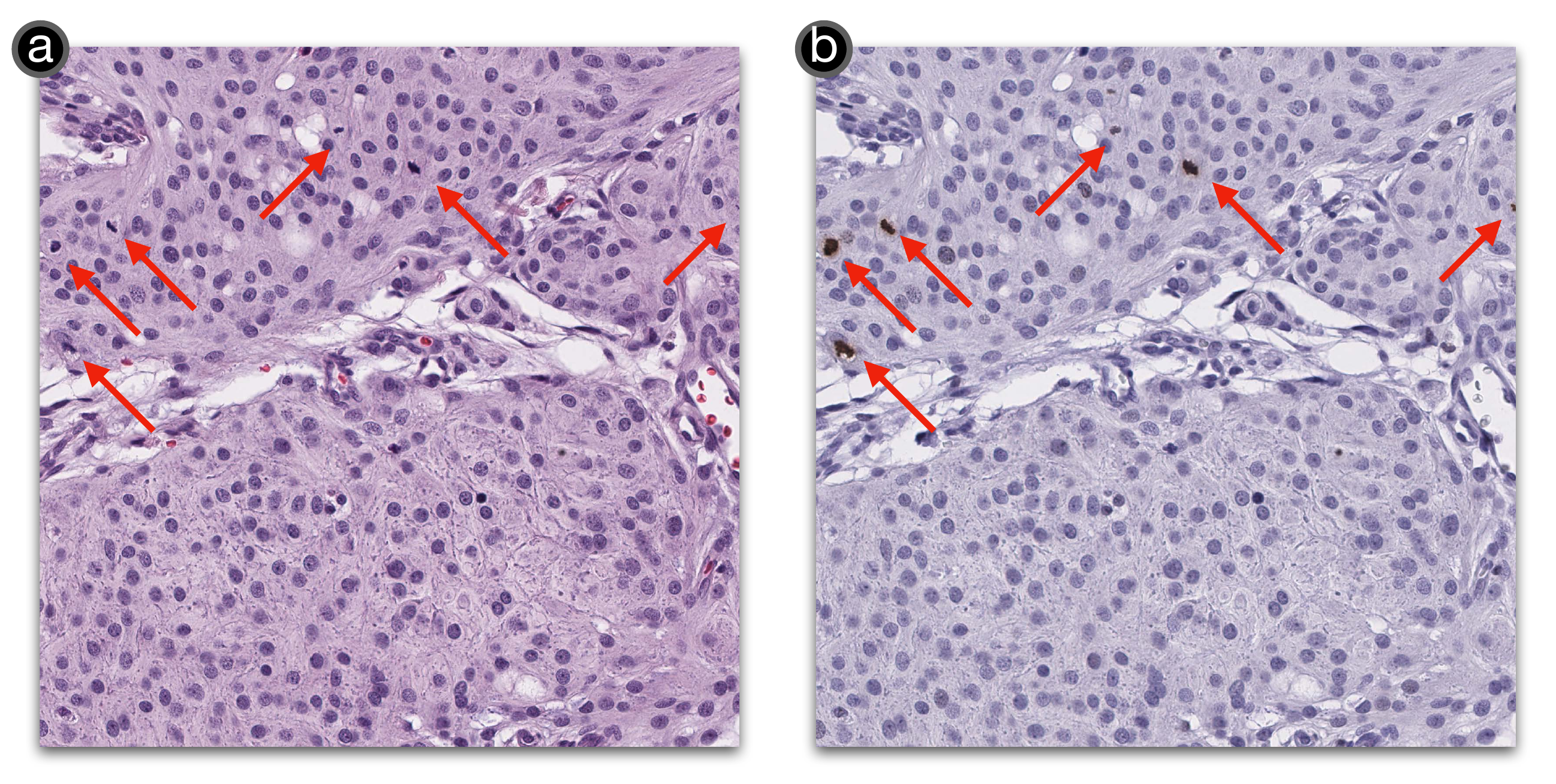}
    \caption{(a) An example region-of-interest image used in the user study, with arrows pointing at the ground truth mitoses; (b) The anti-body test used by the three doctors to annotate the ground truth mitoses. Mitoses were shown in brown (as pointed by the arrows) in the anti-body test.}
    \label{fig:ex_mitos}
\end{figure}

This work selects the task of mitosis (a type of histology pattern) detection in brain tumors of meningiomas (Figure \ref{fig:ex_mitos}(a)). The significance of mitosis stems from its critical role in tumor assessment and patient management for meningiomas \citep{cree_counting_2021, 10.1093/neuonc/noab106, 10.1093/neuonc/noab150}. Despite their importance, pathologists' evaluation of mitoses often faces substantial difficulties. The intricacies lie in mitotic figures' small size, low prevalence, and heterogeneous distribution \citep{aubreville_completely_2020, doi:10.1177/0300985819890686}. These complexities contribute to low reported sensitivities, consistencies among pathologists, and examination efficiencies for mitosis evaluation \citep{collan_standardized_1996, meyer_breast_2005, veta_mitosis_2016, 10.1145/3544548.3580694}, which could negatively impact medical outcomes.

According to the 2021 World Health Organization central nervous system tumor classification guidelines, mitosis serves as a critical diagnostic criterion for grading numerous brain tumors, such as IDH-mutant astrocytoma, oligodendroglioma, and ependymoma \citep{10.1093/neuonc/noab106}. Going beyond mitoses, pathologists may also be required to detect small-scale, sparsely distributed patterns in large scans, such as finding small tumor deposits within lymph nodes in breast cancer or malignant melanoma \citep{regitnig_expectations_2020}. In a more general context, similar visual search tasks also exist in high-stakes domains where AI assistance could be valuable. For instance, security personnel must swiftly identify potential threats like explosives in X-ray scans \citep{wolfe2007low}, and emergency responders rely on timely assessments of disaster impacts from satellite imagery \citep{10.1145/3579481}.

\subsection{Sample Selection \& Mitosis Ground Truth Acquisition}

Meningioma specimens were collected from a local hospital after receiving ethics approval. These specimens were digitized into 19 digital slides with an Aperio CS2 Scanner (Manufacture: Leica, Germany). A specialist pathologist examined these slides and selected 51 regions of interest (ROIs) based on predefined criteria. Each ROI has a dimension of $1,600\times 1,600$ pixels ($400\times 400\mu$m), with one example shown in Figure \ref{fig:ex_mitos}(a). This image dimension matches the field-of-view under the $40\times$ objective lens in light microscopy, which can reduce the mental effort for pathologists to adapt to the digital interface.

As for collecting the mitosis ground truth, two residents independently annotated all 51 images initially. Next, a third specialist pathologist reviewed these initial annotations and provided a final decision. To ensure the accuracy of the ground truth, the three doctors referred to the results of an additional antibody test (the Phosphohistone-H3 immunohistochemistry test, a mitosis indicator usually used in medical research \citep{10.1093/neuonc/nov002, fukushima2009sensitivity}, Figure \ref{fig:ex_mitos}(b)) in the ground truth annotation process.

Within the 51 selected ROI images, three were selected for the tutorial, leaving the rest 48 for testing purposes. The 48 test images have 88 mitoses in total. The count of mitoses per image varies between zero and six, which can cover the majority of mitosis prevalence in a single ROI in meningiomas.

\subsection{Experience Level of Pathologists}

In the United States, pathology professionals can be classified into four levels based on their training progress and experience \citep{genzen2013overview}:

\begin{enumerate}
    \item A \textbf{medical student} is currently receiving medical education.
    \item A \textbf{resident} has earned their Medical Doctor or an equivalent degree and is in post-graduate residency training.
    \item A \textbf{general pathologist} has completed their residency training and holds general board certification in pathology.
    \item A \textbf{specialist pathologist} has received/ is undergoing further training in a sub-specialty area (in this study, neuropathology) after becoming certified as a general pathologist.
\end{enumerate}

Regarding familiarity with the mitosis detection task, specialist pathologists are expected to have the highest level because of their sub-specialty training. General pathologists should have a moderate familiarity, having acquired their general board certification. As for residents and medical students, their familiarity depends on their exposure during rotations and any subsequent training they have received. In this study, 32 medical professionals from ten institutions participated, covering all four aforementioned categories.

% User Study
\section{User Study}

\begin{figure*}
    \centering
    \includegraphics[width=1.0\linewidth]{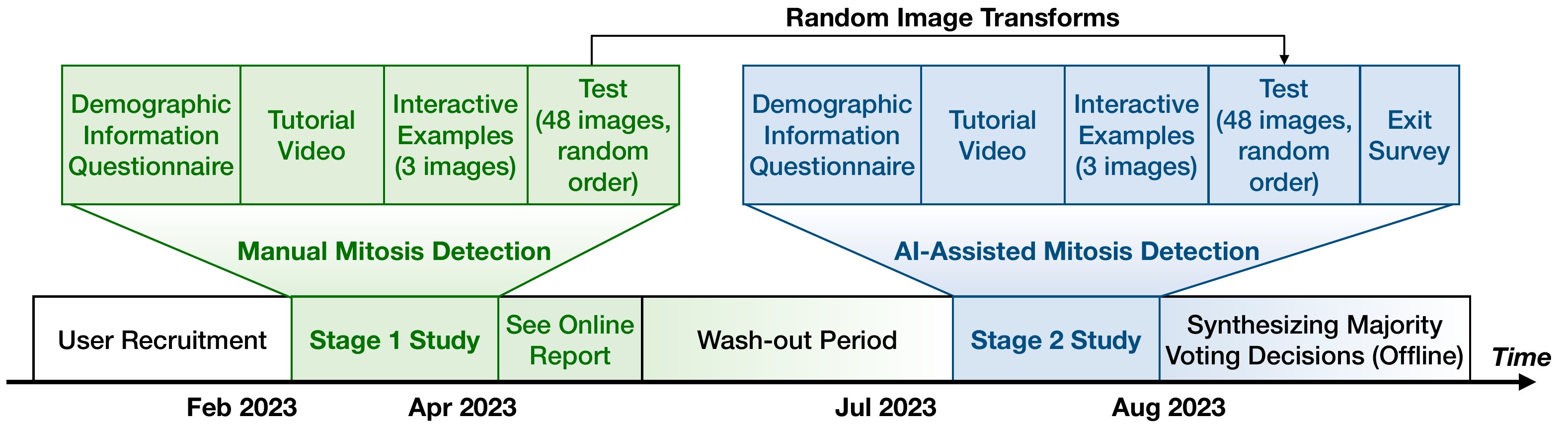}
    \caption{Organization of the user study.}
    \label{fig-study-proc}
\end{figure*}

An online user study was conducted under the Institutional Review Board approval of the University of California, Los Angeles (IRB\#21-000139). The user study has two major stages (Figure \ref{fig-study-proc}): \one \textbf{Stage 1} (February 2023 -- April 2023): participants performed the mitosis detection task in 48 test images manually; \two \textbf{Stage 2} (July 2023 -- August 2023): participants detected mitoses in the same 48 images with AI-assistance. This sequential arrangement follows previous work \citep{10.1145/3581641.3584066},  and was designed to investigate potential shifts in pathologists' decisions influenced by AI. The majority voting decisions were synthesized offline after the stage 2 responses had been collected. The main research questions are:

\begin{itemize}
    \item \textbf{RQ1}: How did pathologists use AI and XAI while performing the ``mitosis detection'' task?
    \item \textbf{RQ2:} How does the majority voting mechanism influence the appropriateness of AI reliance compared to one pathologist collaborating with AI?
    \item \textbf{RQ3}: Is the majority voting mechanism more likely to achieve complementary team performance compared to one pathologist collaborating with AI?
\end{itemize}

\subsection{Participants}
\label{s2p1}
Participants were recruited through sending emails to the mailing list and snowball recruitment. As a result, 32 pathology professionals from 10 medical centers in the United States took part in both study stages, including 12 specialist pathologists, six general pathologists, ten residents, and four medical students\footnote{The four medical student participants underwent a 45-minute training session overseen by a specialist pathologist before participating, to ensure their familiarity with the mitosis detection task.}. The demographic information of participants is shown in the supplemental material.

\subsection{Study Procedure}

Stages 1 and 2 of the user study were conducted in an unmoderated manner. At each stage, each participant joined online with their computers at the recommended display settings. The study of each stage consisted of the following parts (Figure \ref{fig-study-proc}):

\begin{enumerate}
    \item \textbf{Demographic information}: Participants filled in a demographic information questionnaire.
    \item \textbf{Tutorial}: Participants saw a tutorial video describing how to participate, followed by an interactive tutorial of three example images. No AI details were revealed to participants.
    \item \textbf{Test}: Participants examined the 48 images without (stage 1) or with (stage 2) AI assistance. Their task was to detect and report mitoses from these images with their threshold of daily practice.
\end{enumerate}

\begin{figure*}
    \centering
    \includegraphics[width=1.0\linewidth]{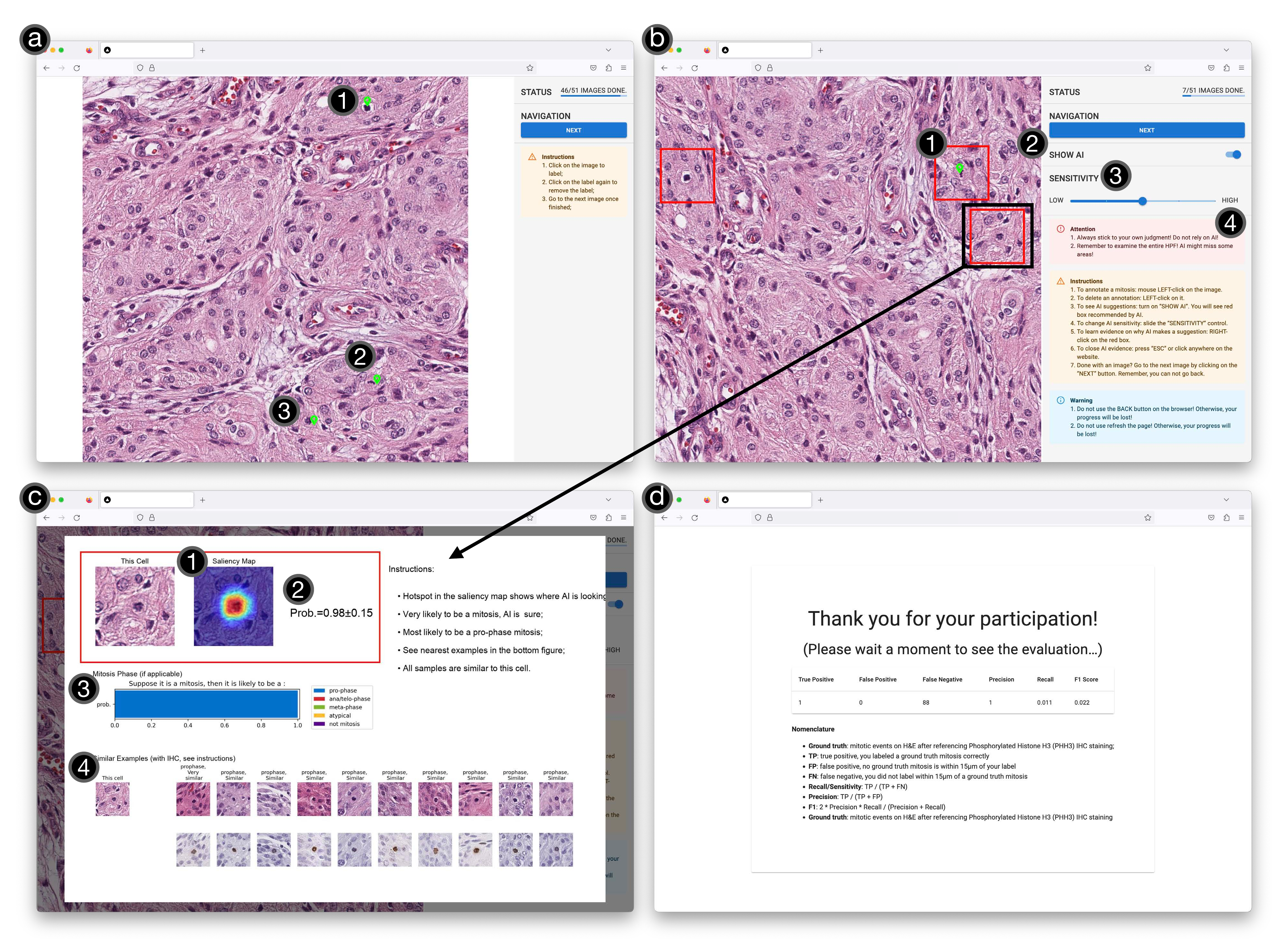}
    \caption{Screenshots of the mitosis study websites: (a) The manual mitosis detection website in the stage 1 study. The user could left-click on the image to leave a mark for each mitosis detected (\textcircled{1} -- \textcircled{3}). (b) The AI-assisted mitosis detection website in the stage 2 study. The interface added \textcircled{1} the AI recommendation box; \textcircled{2} ``Show AI'' switch, where the user could toggle on/off AI recommendations; \textcircled{3} ``AI Sensitivity'' slider, where the user could adjust the sensitivity of AI based on their preference; \textcircled{4} a warning message to remind users not relying on AI. (c) The website in stage 2 also provided an XAI evidence card for each AI recommendation. Each XAI evidence card included \textcircled{1} a saliency map; \textcircled{2} confidence level, including a probability score and a trust score; \textcircled{3} a bar plot for subclass probability; and \textcircled{4} similar examples. (d) After the user finishes examining all images, an evaluation page will inform the performance metrics to the participant.}
    \label{fig:mg-interface}
\end{figure*}

Two methods were introduced to reduce the learning effect of participants in the stage 2:

\begin{itemize}
    \item \textbf{Random image transforms}: Including random flipping (vertical and/or horizontal) and random rotation (randomly chosen from \{0$^{\circ}$, 90$^{\circ}$, 180$^{\circ}$, and 270$^{\circ}$\}). For instance, the image shown in Figure \ref{fig:mg-interface}(b) was rotated 270$^{\circ}$ anti-clockwise from that in Figure \ref{fig:mg-interface}(a). 
    \item \textbf{Wash-out period and ground truth blinding}: After completing stage 1, participants received personalized online report documents highlighting disagreements between their mitosis reportings and the ground truth. After two weeks, they were prevented from accessing these online documents. Next, after a wash-out period of three months, they were invited to participate in the stage 2 study.
\end{itemize}

\subsection{User Interfaces \& Key Features}

For each stage, we deployed an interface online to enable participants to examine the images and report mitoses.

\subsubsection{Stage 1: Manual Mitosis Detection} This interface only showed participants the images and logged their interactions (Figure \ref{fig:mg-interface}(a)). If the user found a mitosis, they could left-click on where it resided to leave a mark (Figure \ref{fig:mg-interface}(a) \textcircled{1} -- \textcircled{3}). The user could go to the next image after examining one. However, they could not return to the previous image to ensure a precise measurement for time consumption. After all images were examined, a status page (Figure \ref{fig:mg-interface}(d)) was displayed to inform the participant of the performance of their mitosis detection.

\subsubsection{Stage 2: AI-Assisted Mitosis Detection} The AI model used in this stage was an EfficientNet-b3 Convolutional Neural Network (CNN), trained from a meningioma mitosis dataset \citep{tan2019efficientnet, 10.1007/978-3-031-33658-4_21}. The website displayed AI mitosis detections through recommendation boxes (Figure \ref{fig:mg-interface}(b)). Additionally, following previous works, we included four components to mitigate the negative influence of improper AI reliance:

\begin{itemize}
    \item \textbf{Warning messages}: A ``black-box'' style\footnote{... the highest safety-related warning assigned by the U.S. Food and Drug Administration \citep{delong2019black}.} warning message was presented in the tutorial video, suggesting that the users should always rely on their judgments. The message was also shown in a highlighted box on the website (Figure \ref{fig:mg-interface}(b) \textcircled{4}).
    
    \item \textbf{XAI}:  Each AI recommendation was accompanied by an evidence card which attempts to provide XAI assistance \citep{plass_explainability_2023}. The user could right-click on the AI recommendation box to see the XAI evidence card \textit{on-demand}. Four popular XAI techniques were included following previous work \citep{EVANS2022281}, including:
    
    \begin{itemize}
        \item \textbf{Saliency map}: Generated by GradCAM++ \citep{gradcam++}.
        \item \textbf{Confidence level}: Including a probability score and a trust score \citep{10.1145/3351095.3372852}. The trust score was the geometric mean of noise \citep{ayhan2018test} and random AI variances \citep{gal2016dropout} of the AI prediction.
        \item \textbf{Subclass}: A bar plot showcasing potential subclasses of the mitosis (\ie pro-phase, meta-phase, ana/telo-phase, atypical, and not mitosis) in this AI recommendation.
        \item \textbf{Similar examples}: A set of ten similar instances was retrieved from an annotated dataset that includes paired Hematoxylin and Eosin -- immunohistochemistry staining \citep{cai2019human}.
    \end{itemize}
        
    Counterfactual explanations were not used because of the low quality of the retrieval results achieved by the our AI model.
    
    \item \textbf{Personalized AI adjustments}: The user could toggle on/off AI recommendations by interacting with the ``Show AI'' switch (Figure \ref{fig:mg-interface}(b) \textcircled{2}) (\ie AI on-request, suggested by \cite{gaube2021ai}) and adjust the AI sensitivity (Figure \ref{fig:mg-interface}(b) \textcircled{3}) according to their preferences. The website provided five AI sensitivity settings for users: ``lowest,'' `low,'' ``medium,'' ``high,'' and ``highest.'' A higher sensitivity would include more AI recommendations with lower probabilities.
    
    \item \textbf{Random image order}: The 48 images were presented to participants in a random order to prevent users from anchoring on AI based on their initial impressions (\ie the ordering effect \citep{10.1145/3397481.3450639}).
\end{itemize}

We chose not to reveal AI information to participants because of the time-consuming nature of the education process.

\subsection{Synthesizing Majority Voting Decisions from Groups of AI-Assisted Participants}
\label{sec44}
\begin{figure*}
    \centering
    \includegraphics[width=1.0\linewidth]{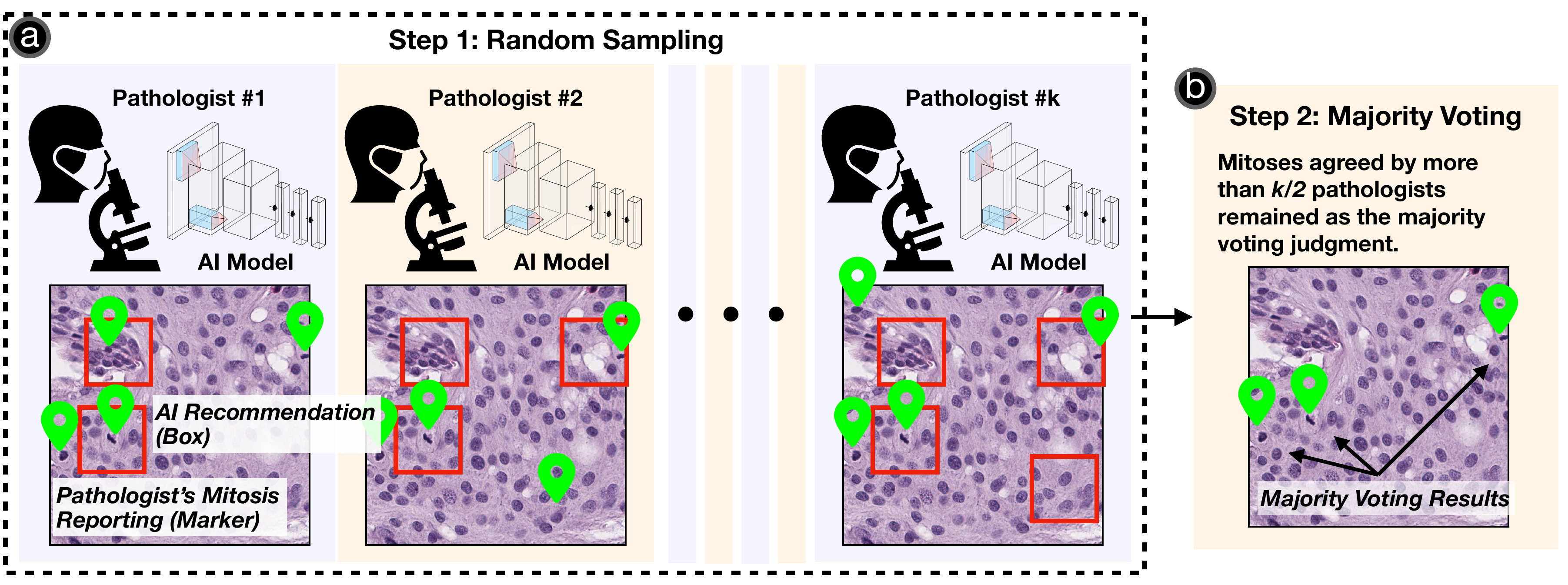}
    \caption{Steps for synthesizing the majority voting decisions from $k$ AI-assisted pathologists: (a) random sampling: mitosis reportings from an odd number of $k$ randomly-sampled, AI-assisted pathologists were collected, (b) majority voting: mitoses candidates reported by $>k/2$ pathologists remained as the final decision.}
    \label{fig-syn-con}
\end{figure*}

Participants' majority voting decisions were synthesized offline after collecting their responses from the stage 2 study. It consisted of two steps:

\begin{itemize}
    \item[\text{Step 1}] \textbf{Random Sampling}: Mitosis reportings from an odd number $k$ participants from stage 2 were aggregated as a group (Figure \ref{fig-syn-con}(a)). Members in a group were sampled randomly from the participant pool without replacement.
    \item[\text{Step 2}] \textbf{Majority Voting}: Mitoses candidates reported by more than half of members ($k/2$) in the group remained as the final majority voting decision (Figure \ref{fig-syn-con}(b)).
\end{itemize}

Group sizes of odd numbers $k =3,5,7, \dots, 27$ were explored. For each group size, the random sampling--majority voting processes were run 100 times for further analysis.

\subsection{Measures \& Statistics}

\subsubsection{Utilization of AI \& XAI (RQ1)}

We employed two metrics to measure how participants used AI assistance in the stage 2 study:
\begin{itemize}
    \item \textbf{AI activation rate}: Indicating the percentage of the 48 test images where the AI was activated at least once (Equation \ref{eq-aia}).
    \item \textbf{AI active time percentage}: Since the participant might deactivate the ``Show AI'' feature, this metric represents the percentage of time when the ``Show AI'' feature stayed active during the entire stage 2 study (Equation \ref{eq-aiatp}).
\end{itemize}

\begin{equation}
\label{eq-aia}
    \text{AI activation rate} = \frac{\sum_{i=1}^{48} \mathds{1}[\text{``Show AI'' in image}_i == \text{``On''}]}{48} \times 100\%
\end{equation}

\begin{equation}
\label{eq-aiatp}
    \text{AI activation time percentage} = \frac{\sum_{i=1}^{48} T[\text{``Show AI'' in image}_i == \text{``On''}]}{\sum_{i=1}^{48} \text{Time consumption on image}_i} \times 100\%
\end{equation}

Participants' utilization of XAI was measured by the following two metrics:
\begin{itemize}
    \item \textbf{XAI activation rate} was calculated according to Equation \ref{eq-xaia}. The number of ``AI recommendations in image$_i$'' was counted based on the highest sensitivity set by a participant while they examined the image$_i$. If the ``Show AI'' was not toggled on in an image, then it was not counted.
    \item \textbf{XAI activation time} was measured by the time elapsed between a participant opening and closing an XAI evidence card.
\end{itemize}

\begin{equation}
\label{eq-xaia}
    \text{XAI activation rate} = \frac{\sum_{i=1}^{48} |\text{XAI opened in image}_i|}{\sum_{i=1}^{48} |\text{AI recommendations in image}_i| \times \mathds{1}[\text{``Show AI'' == ``On''}]} \times 100\%
\end{equation}

\subsubsection{Reliance on AI (RQ2)}
\label{sec454}
\begin{figure*}
    \centering
    \includegraphics[width=1.0\linewidth]{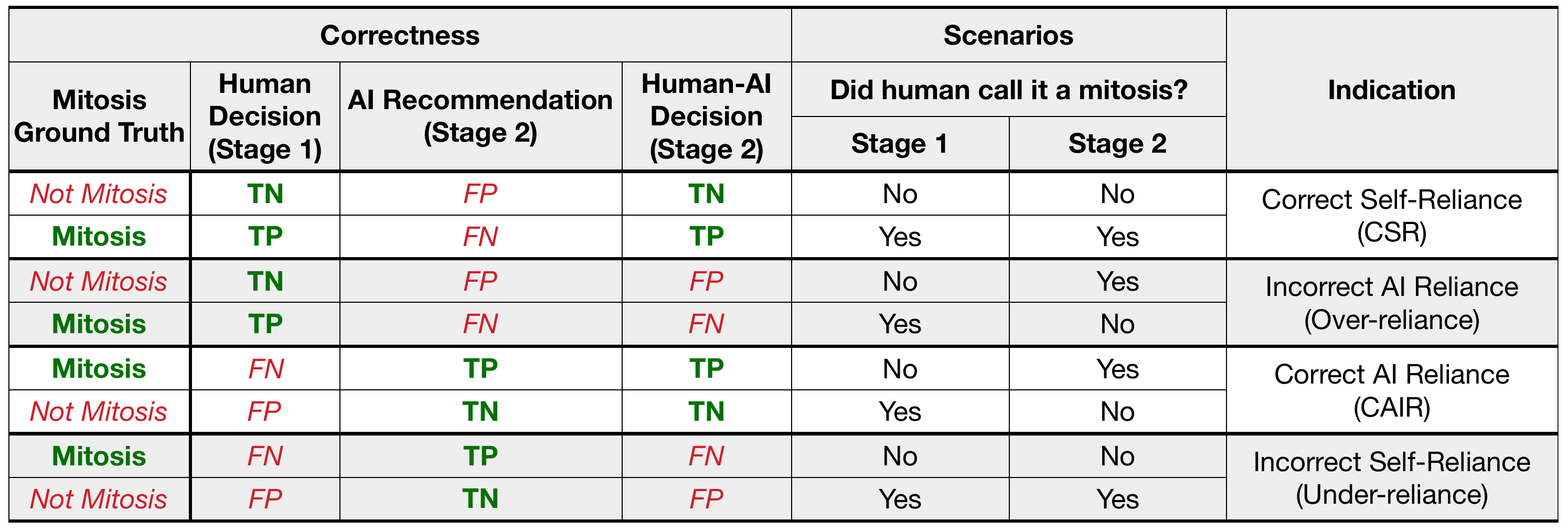}
    \caption{Combinatorics for reliance incidents in the condition of one pathologist collaborating with AI (\ie one-human-AI) for the mitosis detection task. This chart is adopted from the framework described in \cite{10.1145/3581641.3584066}.}
    \label{fig:ai-ri-framework}
\end{figure*}

We used the categorization proposed by \cite{10.1145/3581641.3584066} to define the incidents related to the reliance. Four types of events were defined under the categorization: \one correct self-reliance, \two incorrect AI reliance (over-reliance), \three correct AI reliance, and \four incorrect self-reliance (under-reliance). The criteria for judging these events were based on the true-positive (TP), true-negative (TN), false-positive (FP), and false-negative (FN) detecitons\footnote{A TP was defined as ``there was a ground truth within 60 pixels (15$\mu m$) of a participant-reported mitosis,'' a TN was ``no participant-reported mitoses were found surrounding a non-mitotic figure,'' an FP was ``no ground truth was found within a 60-pixel radius of a participant-reported mitosis,'' and an FN was ``no participant-reported mitoses were found within 60-pixel radius of a ground truth.''}. We adopted the framework in \cite{10.1145/3581641.3584066} for the mitosis detection task, which is summarized in Figure \ref{fig:ai-ri-framework}.

Schemmer \etal further introduced two normalized metrics, Relative AI Reliance (RAIR), and Relative Self-Reliance (RSR), to represent the Appropriateness of Reliance (AoR). The RAIR relates to the under-reliance events (Eq. \ref{eq4}). And the RSR relates to the over-reliance events (Eq. \ref{eq5}). The Appropriateness of Reliance is encapsulated by the tuple of PAIR and RSR (Eq. \ref{eq6}), which can be graphically represented on a 2D chart with the RAIR on the x-axis and the RSR on the y-axis.

\begin{equation}
\label{eq4}
    \text{Relative~AI~reliance~(RAIR)} = %\\
    \dfrac{\text{Correct~AI~Reliance}}{\text{Correct~AI~Reliance} + \text{Under-reliance}}
\end{equation}

\begin{equation}
\label{eq5}
    \text{Relative~Self~reliance~(RSR)} = %\\
    \dfrac{\text{Correct~Self~Reliance}}{\text{Correct~Self~Reliance} + \text{Over-reliance}}
\end{equation}

\begin{equation}
\label{eq6}
    \text{Appropriateness~of~Reliance~(AoR)} = (RSR; RAIR)
\end{equation}

To measure AI reliance on majority voting decisions, we also implemented the majority voting process for stage 1. To ensure a ``with-in-subject'' nature of the analysis, for each majority voting run for stage 2, a vis-\`a-vis majority voting from the same group of participants in stage 1 was conducted. The definitions of ``human decisions,'' ``AI recommendations,'' and ``human-AI decisions'' were adjusted to fit the majority voting condition and are summarized in Table \ref{tab:adj-labels}.

Because participants might employ different AI sensitivity settings in stage 2, the random sampling process to formulate groups was also adopted with regard to each participant's AI sensitivity setting: for the AI reliance analysis, the $k$ pathologists were exclusively drawn from the subset of pathologists who majorly set the same AI sensitivity, which ensured the AI conditions among all group members were similar.

\begin{table*}
    \centering
    \caption{Modified definitions to measure AI reliance for the majority voting decisions synthesized from a group of $k$ pathologists.}
    \begin{tabular}{@{}p{6cm}@{}@{}p{7.5cm}@{}}
    \toprule
        Items &  Majority Voting Decisions (Group Size=$k$)\\
    \midrule
         Human Decision (stage 1) & Majority voting results based on the stage 1 decisions from $k$ participants \\
         AI Recommendation (stage 2) & For each image, AI recommendations under the highest sensitivity set by more than $k/2$ of participants while they were seeing the ROI \\
         Human-AI Decision (stage 2) & Majority voting results based on the stage 2 decisions from the same $k$ participants \\
    \bottomrule
    \end{tabular}
    \label{tab:adj-labels}
\end{table*}

To study \textbf{RQ2}, we compared five conditions: one pathologist collaborating with AI (\ie one-human-AI collaboration), and majority voting for the four group sizes ($k$=3,5,7,9). For each criterion of RAIR and RSR, a Kruskal–Wallis test was first applied to show significance among these five conditions. A post-hoc Dunn's test with Bonferroni correction was then used to test pair-wise significance. Appropriateness of Reliance scatter plots was also drawn to visualize the distribution of RAIR and RSR for these five conditions.

\subsubsection{Correctness of Mitosis Detection (RQ3)}
\label{sec451}
We used \textit{precision} (Eq. \ref{eq1}) and \textit{recall} (Eq. \ref{eq2}) to measure the correctness of the mitosis detection.

\begin{equation}
\label{eq1}
    Precision = \frac{TP}{TP + FP}
\end{equation}

\begin{equation}
\label{eq2}
    Recall = \frac{TP}{TP + FN}
\end{equation}

Here, we compare the precision and recall of five conditions: one-human-AI collaboration, and majority voting decisions from AI-assisted pathologists (group sizes $k$=3,5,7,9). Results of larger group sizes are reported in the supplemental material. Similar to the comparisons in the AI reliance metrics, for each of precision and recall, a Kruskal–Wallis and a follow-up post-hoc Dunn's test with Bonferroni correction was employed to test the significance among the condition pairs.

Because our previous work showed AI achieved higher overall performance than all participants in stage 1 \citep{gu_enhancing_2024}, the ``complementary team performance'' in this work refers explicitly to cases where the human+AI approach outperforms AI (\textbf{RQ3}, \ie super-AI performance). Here, the AI operating point was selected based on the best threshold in the model validation process. For precision and recall, we defined the ``success rate of achieving super-AI performance'' using Equation \ref{eqq4}. This equation was applied to both the one-human-AI collaboration, and majority voting decision conditions with group sizes $k$ ranging from 3 to 27.

\begin{equation}
\label{eqq4}
    \text{Success Rate} = \dfrac{\text{Number of participants/runs exceeding AI performance}}{\text{Total number of participants/runs}} \times 100\%
\end{equation}

% Result
\section{Result}

3/32 participants in stage 2 chose not to activate AI recommendations at all for over 45/48 test images. Therefore, they were classified as non-AI users and were excluded from subsequent analyses. For the remaining 29 participants, we report the utilization of AI and XAI in Section \ref{secr1}. 25/29 of the participants majorly set the sensitivity as either ``highest'' (N=15) or ``medium'' (N=10) during the stage 2 study, and they were included in the AI reliance analysis (Section \ref{secr3}). The responses from all 29 AI-users were used for correctness analyses in Section \ref{secr2}.

\begin{figure*}
    \centering
    \includegraphics[width=1.0\linewidth]{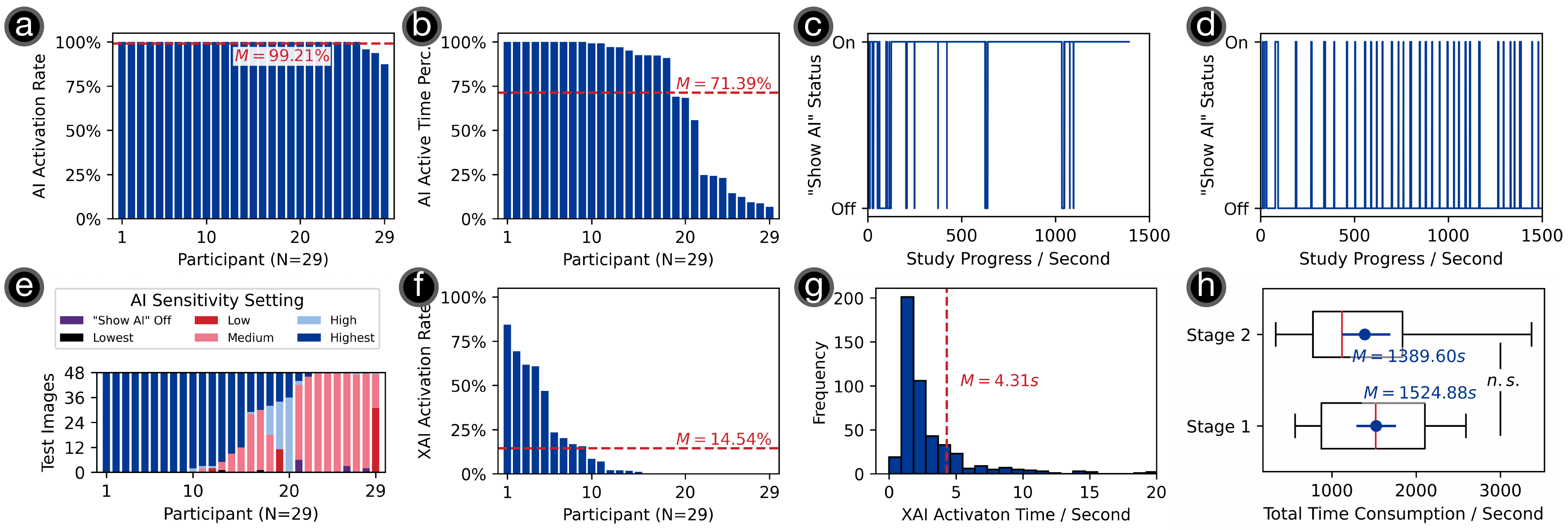}
    \caption{(a) Bar-plot of AI activation rates; (b) Bar-plot of AI active time percentage; Example plots showing how ``Show AI'' status changed for (c) a participant with a high (92.31\%) AI active time percentage and (d) a participant with a low (14.48\%) AI active time percentage; (e) Stacked bar-plot of participants' AI sensitivity settings; (f) XAI activation rates; (g) Histogram of XAI activation time; (h) Box-whisker plot of total time consumption of each participant spent on image examination in the stage 1 and stage 2 study. No significance (n.s.) was observed between the two stages.} 
    \label{fig-r1}
\end{figure*}

\subsection{Utilization of AI \& XAI}
\label{secr1}

The mean AI activation rate was $M=99.21\%$ ($SD=0.481\%$, $CI_{95}=[98.13\%, 100.00\%]$, Figure \ref{fig-r1}(a))\footnote{The mean ($M$), standard deviation ($SD$), and 95\% confidence intervals ($CI_{95}$) were calculated by the bootstrapping method (100\% re-sampling with replacement, 10,000 times) }. And the mean AI active time percentage was $M=71.39\%$ ($SD=6.713\%$, $CI_{95}=[57.75\%, 83.94\%]$, Figure \ref{fig-r1}(b)). 21/29 participants had $>50\%$ AI active time percentages, with an example of how they interacted with the ``Show AI'' feature shown in Figure \ref{fig-r1}(c), which suggests the user kept the AI activated for the majority of the time, with occasional brief flickering between turning it off and on during the initial interactions. The remaining 8/29 participants had $<25\%$ AI active time percentages: Although the ``Show AI'' feature was majority deactivated, these participants would still activate AI recommendations briefly while examining each image (Figure \ref{fig-r1}(d)). Interestingly, this pattern matches the cognitive forcing function \citep{10.1145/3449287} although these participants had not been instructed to do so.

For AI sensitivity settings, 15/29 participants set for the ``highest'' for over half of the ROI images. The remaining participants preferred to set the AI sensitivity as ``high'' (1/29), ``medium'' (10/29), ``low'' (1/29), or showed no clear preference (2/29), as shown in Figure \ref{fig-r1}(e).

Regarding XAI utilization, the mean XAI activation rate was $M=14.54\%$ ($SD=4.537\%$, $CI_{95}=[6.43\%, 24.25\%]$, Figure \ref{fig-r1}(f)). Specifically, 4/29 participants had XAI activation rates higher than 50\%, while 14/29 participants did not activate any XAI at all. The mean XAI activation time was $M=4.31$ seconds ($SD=0.719~\text{seconds}$, $CI_{95}=[3.17~\text{seconds}, 5.95~\text{seconds}]$, Figure \ref{fig-r1}(g)).

On average, participants spent 25 minutes and 25 seconds examining all 48 test images in stage 1, and 23 minutes and 9 seconds in stage 2 (Figure \ref{fig-r1}(h)). The total time consumption did not show a significant difference between the two stages (Wilcoxon rank-sum test, $p=0.31$).

\begin{figure*}
    \centering
    \includegraphics[width=1.0\linewidth]{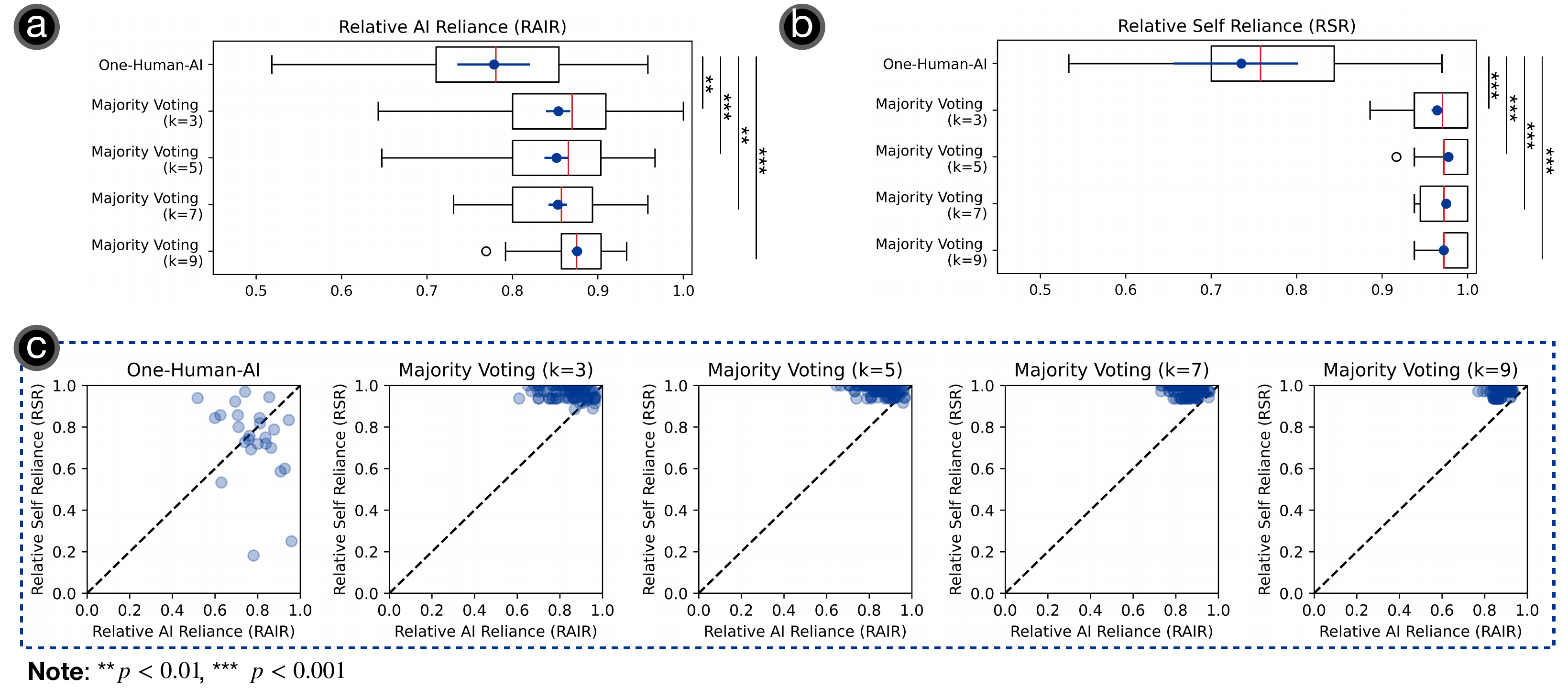}
    \caption{Box-whisker plots of (a) RAIR and (b) RSR for the five conditions of one-human-AI collaboration, and majority voting decisions ($k=3,5,7,9$); (c) Scatter plots for appropriateness of reliance for these five conditions.}
    \label{fig-r2}
\end{figure*}

\subsection{Reliance on AI}
\label{secr3}

As shown in Figure \ref{fig-r2}(a), the mean RAIR of one-human-AI collaboration was $M=0.779$ ($SD=0.021$, $CI_{95}=[0.735, 0.820]$). And that for majority voting decisions of group size $k=3$ was $M=0.852$ ($SD=0.007$, $CI_{95}=[0.839, 0.866]$). The mean RAIR for majority voting decisions of $k=5,7,9$ were 0.866, 0.861, and 0.878. All four majority voting conditions yielded higher RAIR ($\sim 9\%$ increase) than one-human-AI collaboration. A Kruskal–Wallis test showed a significant difference among the RAIR values across five conditions ($\eta^2_H=0.043$, $p<0.001$). Post-hoc Dunn's test with Bonferroni correction indicated significance in comparison pairs of one-human-AI \vs majority voting decision from group sizes of $k=3$ ($p=0.012$), $k=5$ ($p<0.001$), $k=7$ ($p=0.004$), and $k=9$ ($p<0.001$).

The mean RSR of one-human-AI collaboration was $M=0.735$ ($SD=0.037$, $CI_{95}=[0.657, 0.803]$, see Figure \ref{fig-r2}(b)). As a comparison, the mean RSR of majority voting decisions of $k=3$ was $M=0.964$ ($SD=0.003$, $CI_{95}=[0.959, 0.970]$). The RSR of majority voting decisions for $k=5,7,9$ were 0.968, 0.976, and 0.967. Similarly, all four majority voting conditions led to higher RSR ($\sim31\%$ increase). A Kruskal–Wallis test showed a significant difference among the RSR values across five conditions ($\eta^2_H=0.178$, $p<0.001$). Post-hoc Dunn-Bonferroni test showed significance in comparison pairs of one-human-AI \vs majority voting decision from group sizes of $k=3$ ($p<0.001$), $k=5$ ($p<0.001$), $k=7$ ($p<0.001$), and $k=9$ ($p<0.001$).

Figure \ref{fig-r2}(c) presents the appropriateness of reliance (AoR) scatter plots for five conditions. These plots demonstrate that majority voting decisions could improve higher RAIR and RSR simultaneously, indicating a high level of AoR was achieved.

\subsection{Correctness of Mitosis Detection}
\label{secr2}

\begin{figure*}
    \centering
    \includegraphics[width=1.0\linewidth]{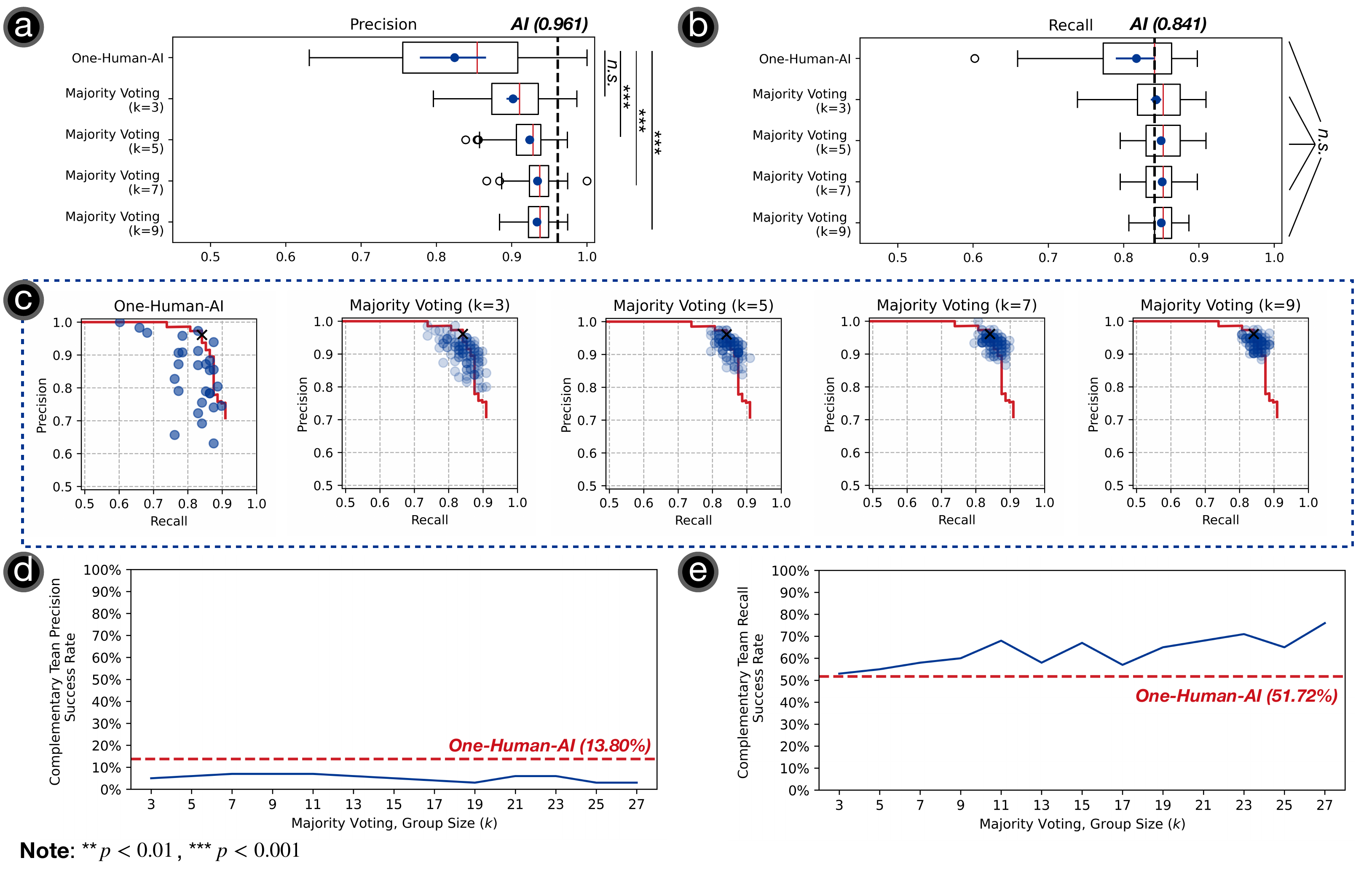}
    \caption{Box-whisker plots of precision and recall for the five conditions of one-human-AI collaboration, and majority voting decisions ($k=3,5,7,9$); (c) Precision-recall plots for mitosis detection for these five conditions. The red line represents the precision-recall curve of AI, and the `x' marker indicates the AI's performance at a threshold determined by the best validation performance. The success rates of achieving super-AI} performance (\ie percentage of human+AI cases where their performance is higher than both humans and AI) for the criteria of (d) precision and (e) recall.
    \label{fig-r3}
\end{figure*}

As shown in Figure \ref{fig-r3}(a), the mean precision of one-human-AI collaboration was $M=0.824$ ($SD=0.023$, $CI_{95}=[0.776, 0.867]$). For majority voting decisions of $k=3$, the mean precision was $M=0.902$ ($SD=0.004$, $CI_{95}=[0.893, 0.910]$). The majority voting of $k=5,7,9$ had mean precisions of 0.924, 0.934, and 0.934, respectively. The AI achieved a higher precision of 0.961. All four majority voting conditions achieved higher precision ($\sim8\%$ increase) than the one-human-AI collaboration. A Kruskal–Wallis test showed that the precision significantly differed across five conditions ($\eta^2_H=0.150$, $p<0.001$). Post-hoc Dunn-Bonferroni test did not observe significance in the comparison pair of one-human-AI \vs majority voting decision $k=3$ ($p=0.715$). Statistical significance was observed for comparison pairs of one-human-AI \vs majority voting decision $k=5$ ($p<0.001$), $k=7$ ($p<0.001$), and $k=9$ ($p<0.001$).

The mean recall of one-human-AI collaboration was $M=0.817$ ($SD=0.013$, $CI_{95}=[0.790, 0.841]$, see Figure \ref{fig-r3}(b)). Majority voting decisions of $k=3$ had a mean recall of $M=0.843$ ($SD=0.003$, $CI_{95}=[0.838, 0.851]$). Majority voting decisions of $k=5,7,9$ had mean precisions of 0.850, 0.851, and 0.850. In comparison, AI achieved a precision of 0.841. Kruskal–Wallis test did not show that the recall differed significantly across five conditions ($\eta^2_H<0.001$, $p=0.774$).

Figure \ref{fig-r3}(c) presents the precision-recall scatter plots for the five conditions. The plots reveal that the majority voting decision exhibits lower variation in both precision and recall compared to the one-human-AI collaboration, indicating a more robust performance. This observation is further supported by the lower $SD$ values for the majority voting decisions, as reported above.

Regarding the success rates for achieving super-AI performance, for precision, none of the majority voting conditions (\ie $k=3 \rightarrow 27$) was higher than the success rate achieved by one-human-AI collaboration ($13.87\%$ success rate, Figure \ref{fig-r3}(d)). On the other hand, for recall, all majority voting conditions had higher success rates compared to one-human-AI collaboration ($51.72\%$ success rate): As shown in Figure \ref{fig-r3}(e), the lowest success rate was observed at $k=3$ ($53\%$ success rate), and the highest was achieved at $k=27$, reaching $76\%$.

% Discussion
\section{Discussion}

\subsection{Summary of Result}

\subsubsection{Summary of RQ1} 

For most participants, AI was activated at least once in most images. However, this does not imply that the AI was constantly active throughout the entire study. Notably, 8/29 participants deactivated AI for most of the study, and only activated it briefly occasionally. That is, in certain instances, the `AI on-request' feature posed cognitive forcing function effects.

The utilization of XAI was relatively low; only four participants opened more than 50\% of the XAI evidence, while nearly half of the participants did not open any. Even when XAI was opened, the time spent by participants on viewing XAI was relatively short (about four seconds) -- in the context of pathologist-AI collaboration, the effectiveness of XAI in mitigating over-reliance may be limited. This is likely because the time-pressing nature of the pathology task outweighed the benefit of XAI explanations, causing pathologists to use XAI less in practice. In light of this, we argue that alternative approaches, such as the majority voting used in this study, need to be investigated to enable appropriate AI reliance for future pathology applications.

\subsubsection{Summary of RQ2}

Pair-wise statistical tests revealed significant improvements in both RAIR and RSR metrics for majority voting decisions ($k=3, 5, 7, 9$), compared to one pathologist collaborating with AI. Specifically, RAIR showed an approximate 9\% increase, and RSR showed about 31\% increase. The PAIR-RSR scatter plots indicated simultaneous improvements in both metrics. Such results demonstrate a reduction in the proportion of over-reliance against correct self-reliance events, and under-reliance against correct AI reliance events, indicating a higher level of appropriateness of reliance was achieved (according to the definitions in \cite{10.1145/3581641.3584066}).

\subsubsection{Summary of RQ3}
No significant difference in the precision was observed between the condition of one-human-AI collaboration and the majority voting with $k=3$. A statistical significance in the precision was observed when increasing $k$ to 5, 7, and 9. The majority voting conditions improved precision by approximately $8\%$. For recall, no significant differences were observed. The precision-recall scatter plots demonstrated that majority voting decisions exhibited lower variation, suggesting that they were robust and less prone to be influenced by the sample selection.

All majority voting conditions for $k=3\rightarrow 27$ did not show a higher success rate in achieving super-AI precision than one-human-AI collaboration. This is because AI had a high precision of 0.961, and there was a lack of space for improvement. For recall, all majority voting conditions ($k=3\rightarrow 27$) showed higher success rates. Notably, the highest success rate, 76\%, was achieved at $k=27$, indicating a 46.95\% increase over the one-human-AI collaboration condition (51.72\% success rate).

\begin{figure*}
    \centering
    \includegraphics[width=1.0\linewidth]{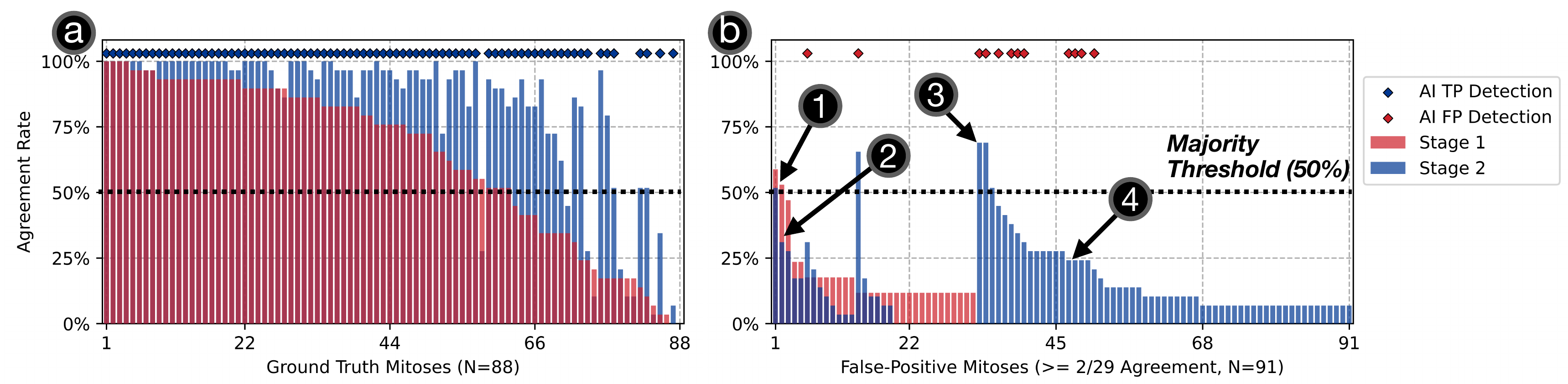}
    \caption{Bar plots for the agreements among 29 participants for (a) 88 ground-truth mitoses, and (b) 91 false-positive mitoses that at least two participants agreed on. The diamond markers ($\lozenge$) stand for the AI detections under the ``Highest'' AI sensitivity setting. \textcircled{1} An example of under-reliance that might not be addressed by the majority voting; \textcircled{2} An example of under-reliance that might be addressed by the majority voting; \textcircled{3} An example of over-reliance that might not be addressed by the majority voting; and \textcircled{4} An example of over-reliance that might be addressed by the majority voting.}
    \label{fig-d1}
\end{figure*}

\subsection{The Mechanism and Cost of Majority Voting}
\label{sec62}
To further explore why the majority voting mechanism was effective, we introduced a metric, ``agreement rate,'' defined as the percentage of participants the reported a cell as a mitosis (regardless of its actual status). We calculated the agreement rates of the 29 participants in both stage 1 and stage 2 studies. These agreement rates covered all 88 ground truth mitoses (Figure \ref{fig-d1}(a)) and 91 false-positive mitoses reported by at least two participants (Figure \ref{fig-d1}(b)). According to Section \ref{sec44}, cells with agreement rates higher than 50\% should be kept as the majority voting decisions. While Figure \ref{fig-d1} is not directly applicable for interpreting results in smaller sub-groups (\eg $k=3$), it illustrates the general trends in participants' agreement rates when influenced by AI. The data revealed two key insights: 

\begin{itemize}
\item \textbf{Reducing Over-Reliance on AI False Positives}: AI's false-positive detections led to higher agreement rates among participants (as shown in Figure \ref{fig-d1}(b)\textcircled{3}), suggesting participants' tendency of over-reliance in at stage 2. The majority of these false-positive detections did not achieve agreement rates higher than 50\% (Figure \ref{fig-d1}(b)\textcircled{4}). In other words, from a group's perspective, it was not usual for the majority of participants to consistently over-rely when AI made false-positive mistakes. Therefore, the over-reliance can be reduced by the majority voting mechanism.
\item \textbf{Reducing Under-Reliance on Human False Positives:} At stage 2, participants may make the same false-positive mistake as in stage 1, even when AI correctly suggested negative (Figure \ref{fig-d1}(b)\textcircled{1}). This suggests that the under-reliance incidents happened when one participant collaborated with AI. Nevertheless, agreement rates for these false positives rarely exceeded the 50\% majority threshold (Figure \ref{fig-d1}(b)\textcircled{2}), indicating that majority voting could reduce under-reliance.
\end{itemize}

\begin{figure}
    \centering
    \includegraphics[width=0.7\linewidth]{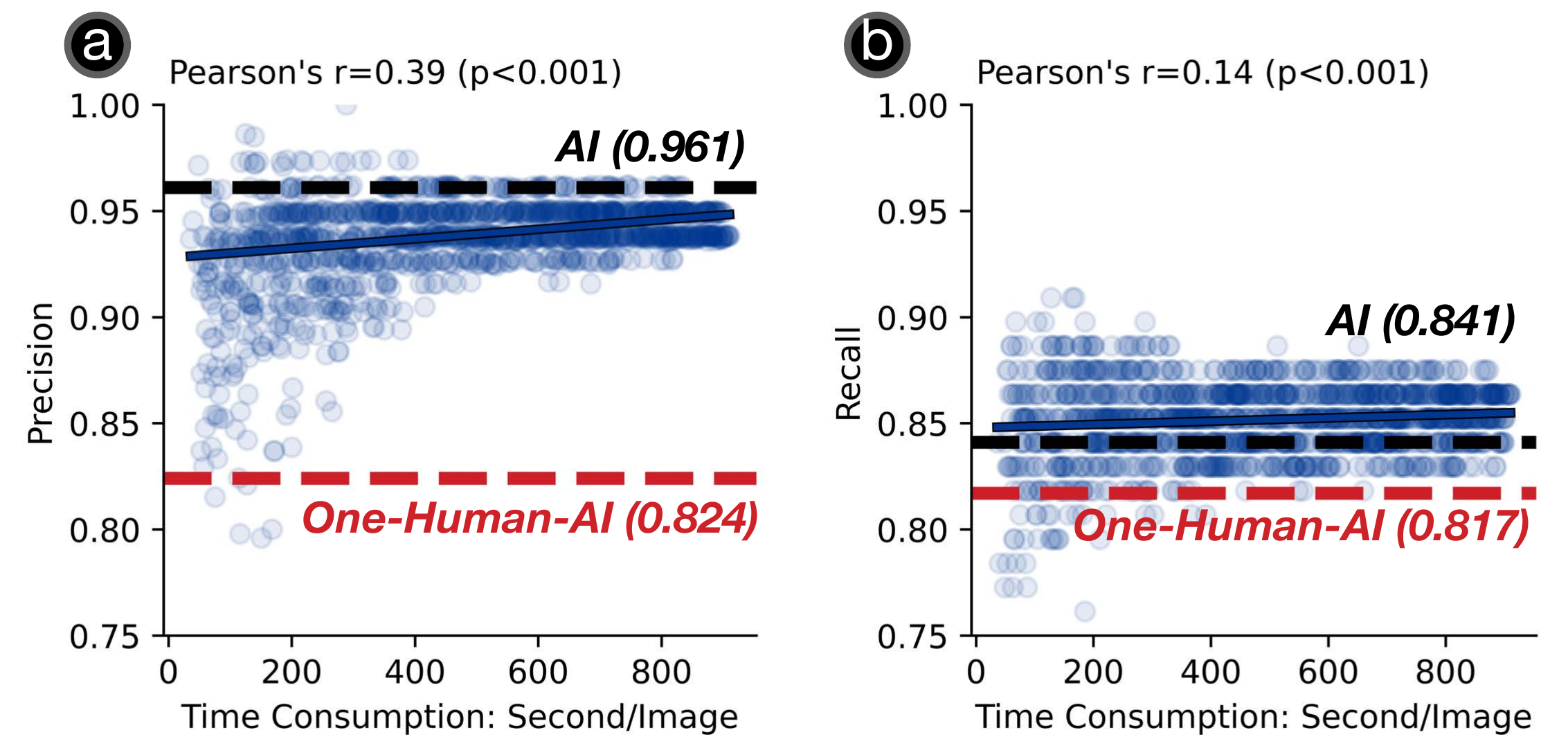}
    \caption{Linear regression plots studying the relations between (a) precision-time consumption, and (b) recall-time consumption while synthesizing majority voting decisions, $k=3\rightarrow 27$, $n=100$ for each $k$.}
    \label{fig-d2}
\end{figure}

To understand the underlying cost of the majority voting mechanism, we analyzed time consumption spent on employing multiple pathologists, and its association with the correctness. Specifically, we conducted 100 majority voting runs for each group size ($k$) ranging from 3 to 27. We applied Pearson's correlation analysis to assess the relationship between precision or recall achieved in each run and its corresponding time consumption. We found a moderate positive correlation between precision and time consumption (Pearson's $r=0.39$, $p<0.001$, $N=1,300$, Figure \ref{fig-d2}(a)), and a weak positive correlation between recall and time consumption (Pearson's $r=0.14$, $p<0.001$, $N=1,300$, Figure \ref{fig-d2}(b)). Certain runs with a relatively small time consumption could reach considerable precision and recall. Note that this is a `bare minimum' estimation: Delays caused by coordinating pathologists should be taken into account in practical applications.

\subsection{On Developing Structured Decision-Making Processes with AI+\textit{k}}

Different from traditional one-human-AI collaboration (AI+1), this study sets the first step towards multiple medical professionals collaborating with AI (AI+$k$) using a simple majority voting technique. We argue that this majority voting approach has three advantages: \one It is flexible and has a simple structure, eliminating the need for face-to-face or online discussions; \two It keeps participants anonymous, thus reducing potential social pressure; \three It is inherently democratic, ensuring that each participant's opinion has an equal weight. We found that this majority voting approach could effectively improve the appropriateness of reliance, and achieve higher-quality medical decisions. As for the limitations of majority voting, one may argue that this approach does not incorporate the discussion process, and decisions with conflicts (\ie $\sim 50\%$ agreement rates) cannot be addressed easily.

Future works might explore AI+$k$ decision-making techniques that involve structured or semi-structured face-to-face discussions \citep{black1999consensus}. Traditionally, these discussions were moderated by the humans. Nonetheless, we envision that future AI can not only help each group member to reach a decision (\eg help pathologists detect mitoses in this study), but can moderate the discussions. For instance, a large language model (LLM) \citep{10.1145/3605943} might anonymously gather and summarize comments from each group member and present a consolidated overview to the group. Members could then have an opportunity to revise their decisions after hearing from the LLM's summary. Given the LLM's omni-availability, no conflict of interest, and and impartiality to authority or personal factors, such AI-facilitated discussions could offer advantages in speed and bias correction, compared to traditional discussion coordination with human moderators.

\subsection{Towards Efficient \& Reliable Medical Decisions with AI+\textit{k}}

Section \ref{sec62} showed that the performance of majority voting decisions from AI+$k$ showed a positive relation to the time consumption. In other words, in general, the more medical professionals involved, the higher the quality of the majority voting decision. Typically, high-risk medical decisions involve 7--10 group members \citep{mcmillan2016use}, while groups as large as 27 done in this study were quite rare. Therefore, considering the time taken to reach a result, we argue that not all medical decisions necessitate the AI+\{\textit{large k}\} approach: cases with high confidence from both AI and humans could be adjudicated by smaller groups with as few as three experts, while those with low AI confidence or prone to human errors could benefit from incrementally larger group sizes, which can yield better and more robust outcomes.

Determining the optimal balance between decision-making and time expenditure has been well-explored in previous crowd-sourcing works \citep{10.1145/3148148}. However, one should be aware that the workflow of medical professionals is usually different from that of general users, and their preferences in using AI and XAI may also vary (as shown in Section \ref{secr1}). Therefore, future research should focus on exploring which AI+$k$ methods can seamlessly integrate into the workflow of medical professionals, effectively balancing efficiency and reliability in the medical decisions of multiple doctors. Additionally, investigating the role of counterfactual explanations \citep{zhou_generating_2022, del_ser_generating_2024} to build trust and facilitate appropriate AI reliance could complement approaches like majority voting, potentially improving interpretability and familiarity with the decision process when integrating AI into risk-sensitive medical workflows.

\subsection{Limitations \& Future Work}

The following points are the limitations of this study and are regarded as future work.

\begin{itemize}
    \item The majority voting synthesizing process did not involve any discussion or communication among participants, which could influence the outcomes.
    \item A 50\% threshold was used to represent the majority. Other thresholds and their impacts were not investigated.
    \item The potential learning effect, particularly among participants in training (\ie residents and medical students), between stage 1 and stage 2 of the study cannot be ignored.
    \item All participants were from one country, potentially limiting the generalizability of findings.
\end{itemize}

% Conclusion
\section{Conclusion}
This study introduces and validates the majority voting approach to enable doctors' appropriate reliance on medical AI. By recruiting 32 pathology professionals, we conducted a multi-institutional, multi-stage user study focusing on detecting mitoses in tumor images. Our analysis revealed that even with groups of three doctors, the majority-voting decisions had a higher appropriateness of AI reliance, compared to one doctor collaborating with AI. Subsequently, the majority voting decisions demonstrated increased precision and recall, although no statistical significance in recall was observed. Additionally, majority voting decisions were more likely to achieve super-AI performance in the recall. While effective on its own, majority voting can also be used together with other techniques to enable appropriate AI reliance. Involving multiple experts in decision-making can yield higher-quality, more robust outcomes that are less prone to AI errors, which holds promise in pathology and broader high-stakes domains.

\bibliographystyle{elsarticle-harv} 
\bibliography{reference}

\newpage
\appendix

\section{Demographic Information of the Participants}
\label{app_demo}
32 participants submitted their responses in both stages of the user study, including 12 specialist pathologists, 6 general pathologists, 10 pathology residents, and 4 medical students. 18/32 participants were from \textbf{Institution \#1}(I1), 5/32 from I2, 2/32 from I3, and the remaining 7/32 were each from a different institution. Their demographic information was summarized in Table \ref{tab:p} (YoE: Years of Experience).

\begin{itemize}
    \item[\textbf{Note 1}] Did not activate AI for over 45/48 images at Stage 2 study. Considered as non-AI users and excluded from the all analyses.
    \item[\textbf{Note 2}] Years of experience (YoE) not applicable for the medical student. To ensure their familiarity with the mitosis detection task, all medical student participants underwent a 45-minute training session overseen by a specialist pathologist before participating.
\end{itemize}

\begin{table}
    \centering
    \caption{Demographic Information of the Participants}
    \begin{tabular}{ccccc}
    \toprule
        \textbf{Index} & \textbf{Experience Level} & \textbf{Institution} & \textbf{YoE} & \textbf{Note} \\
    \midrule
        1 & Specialist Pathologist & I1 & 5--10 & \\
        2 & Specialist Pathologist & I2 & 5--10 & \\
        3 & Specialist Pathologist & I3 & $>$10 & \\
        4 & Specialist Pathologist & I4 & 5--10 & \\
        5 & Specialist Pathologist & I1 & 5--10 & \\
        6 & Specialist Pathologist & I2 & $>$10 & See \textbf{Note 1} \\
        7 & Specialist Pathologist & I5 & $>$10 & \\
        8 & Specialist Pathologist & I6 & $>$10 & \\
        9 &  Specialist Pathologist & I2 & 5--10 & \\
        10 &  Pathology Resident & I1 & 2--5 & \\
        11 &  Specialist Pathologist & I7 & 5--10 & \\
        12 &  Pathology Resident & I2 & 2--5 &\\
        13 &  Pathology Resident & I1 & 2--5 & \\
        14 &  Pathology Resident & I1 & 2--5 & See \textbf{Note 1} \\
        15 &  Specialist Pathologist & I8 & 5--10 & \\
        16 &  General Pathologist & I2 & 5--10 & \\
        17 &  General Pathologist & I9 & 5--10 & \\
        18 &  General Pathologist & I1 & 5--10 & \\
        19 &  General Pathologist & I1 & 5--10 & \\
        20 &  General Pathologist & I1 & $>$10 & See \textbf{Note 1} \\
        21 &  Pathology Resident & I1 & 2--5 &\\
        22 &  Pathology Resident & I3 & 2--5 &\\
        23 &  General Pathologist & I1 & 5--10 & \\
        24 &  Medical Student & I1 & N/A & See \textbf{Note 2} \\
        25 &  Medical Student & I1 & N/A & See \textbf{Note 2} \\
        26 &  Pathology Resident & I1 & 2--5 & \\
        27 &  Specialist Pathologist & I10 & $>$10 & \\
        28 &  Pathology Resident & I1 & 2--5 &\\
        29 &  Pathology Resident & I1 & 2--5 &\\
        30 &  Pathology Resident & I1 & 2--5 & \\
        31 &  Medical Student & I1 & N/A & See \textbf{Note 2} \\
        32 &  Medical Student & I1 & N/A & See \textbf{Note 2} \\
    \bottomrule
    \end{tabular}
    \label{tab:p}
\end{table}

\section{Precision and Recall of Majority Voting Decisions with Larger Group Sizes}
\label{app_pr}
\begin{figure}
    \centering
    \includegraphics[width=1.0\linewidth]{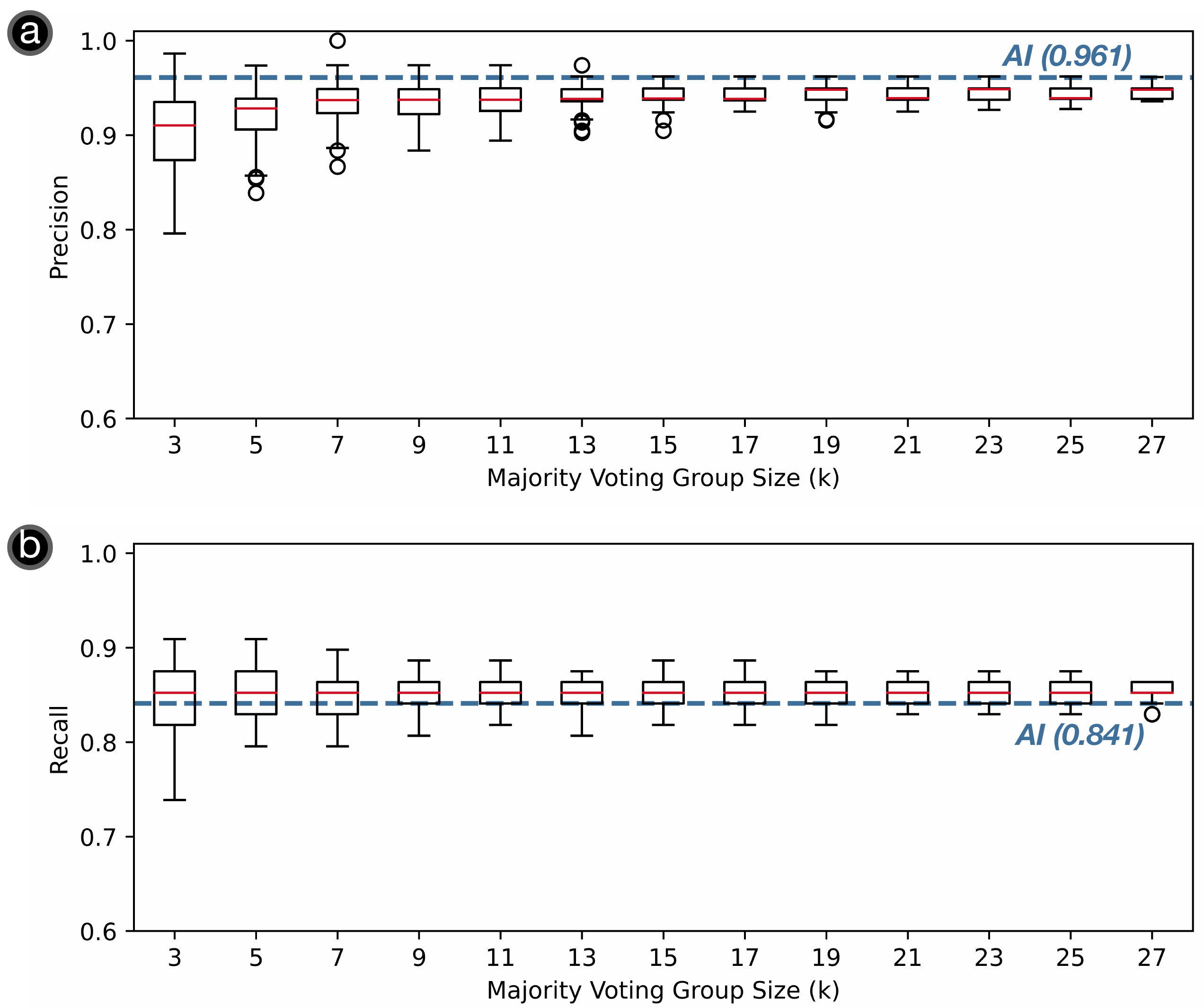}
    \caption{(a) Precision and (b) recall values for the majority voting decisions (AI-assisted, stage 2), group size $k=3\rightarrow27$}
    \label{fig:enter-label}
\end{figure}

\end{document}